\documentclass[letterpaper,twocolumn,prl,aps,superscriptaddress,floatfix]{revtex4}

\usepackage[latin9]{inputenc}
\usepackage{amsmath}
\usepackage{amssymb}
\usepackage{graphicx}
\usepackage{esint}
\usepackage{times}

\usepackage[unicode=true,
bookmarks=true,bookmarksnumbered=false,bookmarksopen=false,
breaklinks=false,pdfborder={0 0 1},backref=false,colorlinks=true]
{hyperref}
\hypersetup{
    linkcolor=magenta, urlcolor=blue, citecolor=blue, pdfstartview={FitH}, hyperfootnotes=false, unicode=true}

\makeatletter

\pdfpageheight\paperheight
\pdfpagewidth\paperwidth

\@ifundefined{textcolor}{}
{%
    \definecolor{BLACK}{gray}{0}
    \definecolor{WHITE}{gray}{1}
    \definecolor{RED}{rgb}{1,0,0}
    \definecolor{GREEN}{rgb}{0,1,0}
    \definecolor{BLUE}{rgb}{0,0,1}
    \definecolor{CYAN}{cmyk}{1,0,0,0}
    \definecolor{MAGENTA}{cmyk}{0,1,0,0}
    \definecolor{YELLOW}{cmyk}{0,0,1,0}
}

\usepackage{xcolor}\usepackage{soul}

\newcommand{\bra}[1]{\ensuremath{\left\langle#1\right|}}
\newcommand{\ket}[1]{\ensuremath{\left|#1\right\rangle}}
\definecolor{blue}{rgb}{0,0,1}
\definecolor{red}{rgb}{1,0,0}
\definecolor{green}{rgb}{0,1,0}

\makeatother

\begin{document}

\title{Experimental implementation of universal nonadiabatic geometric quantum gates in a superconducting circuit}

\author{Y. Xu}
\thanks{These three authors contributed equally to this work.}
\affiliation{Center for Quantum Information, Institute for Interdisciplinary Information
Sciences, Tsinghua University, Beijing 100084, China}
\author{Z. Hua}
\thanks{These three authors contributed equally to this work.}
\affiliation{Center for Quantum Information, Institute for Interdisciplinary Information
Sciences, Tsinghua University, Beijing 100084, China}
\author{Tao Chen}
\thanks{These three authors contributed equally to this work.}
\affiliation{Guangdong Provincial Key Laboratory of Quantum Engineering and Quantum Materials, GPETR Center for Quantum Precision Measurement, and School of Physics and Telecommunication Engineering, South China Normal University, Guangzhou 510006, China}
\author{X. Pan}
\affiliation{Center for Quantum Information, Institute for Interdisciplinary Information
Sciences, Tsinghua University, Beijing 100084, China}
\author{X. Li}
\affiliation{Center for Quantum Information, Institute for Interdisciplinary Information
Sciences, Tsinghua University, Beijing 100084, China}

\author{J. Han}
\author{W. Cai}
\author{Y. Ma}
\author{H. Wang}
\author{Y.P. Song}
\affiliation{Center for Quantum Information, Institute for Interdisciplinary Information Sciences, Tsinghua University, Beijing 100084, China}

\author{Zheng-Yuan Xue} \email{zyxue83@163.com}
\affiliation{Guangdong Provincial Key Laboratory of Quantum Engineering and Quantum Materials, GPETR Center for Quantum Precision Measurement, and School of Physics and Telecommunication Engineering, South China Normal University, Guangzhou 510006, China}

\author{L.~Sun} \email{luyansun@tsinghua.edu.cn}
\affiliation{Center for Quantum Information, Institute for Interdisciplinary Information
Sciences, Tsinghua University, Beijing 100084, China}

\begin{abstract}
Using geometric phases to realize noise-resilient quantum computing is an important method to enhance the control fidelity. In this work, we experimentally realize a universal nonadiabatic geometric quantum gate set in a superconducting qubit chain. We characterize the realized single- and two-qubit geometric gates with both quantum process tomography and randomized benchmarking methods. The measured average fidelities for the single-qubit rotation gates and two-qubit controlled-Z gate are 0.9977(1) and 0.977(9), respectively. Besides, we also experimentally demonstrate the noise-resilient feature of the realized single-qubit geometric gates by comparing their performance with the conventional dynamical gates with different types of errors in the control field. Thus, our experiment proves a way to achieve high-fidelity geometric quantum gates for robust quantum computation.
\end{abstract}
\maketitle
\vskip 0.5cm

\narrowtext

In quantum physics, wave functions up to a global phase are equivalent, and thus the important role played by the phase factors had been ignored for a long time. However, the evolution of a quantum state can be traced in some extent by a geometric phase factor. A famous example is the Aharonov-Bohm effect~\cite{Aharonov1959Significance}, which shows that the phases with a geometric origin can have observable consequences~\cite{berry1984quantal}. Different from the dynamical phase, geometric phases~\cite{berry1984quantal,wilczek1984appearance,Aharonov1987Phase} are gauge invariant and depend only on the global properties of the evolution path. Therefore, besides their fundamental importance, geometric phases have been tested in a variety of settings and have found many interesting applications ~\cite{Wilczek1989Geometric,Xiao2010Berry,Nesterenko2014Transport}.

Recently, there is a renewed interest in applying geometric phases into the field of quantum computation~\cite{Zanardi1990Holonimic, Pachos1999Nonabelian, sjoqvist2008trend}, which is potentially capable to handle hard problems for classical computers~\cite{Nielsen}. The reason is that the global properties of the geometric phases can be naturally used to achieve noise-resilient quantum manipulation against certain local noises~\cite{zhu2005geometric,solinas2012on,johansson2012robustness}, which is essential for practical quantum computation. With adiabatic cyclic evolutions, recent experiments have reported the detection of geometric phases~\cite{adiabatic,adiabatic1,adiabatic2,adiabatic3,Pechal2012Geometric,Berger2013Exploring, Gasparinetti2016Measurement,Yale2016Optical,adiabatic7} and the realization of
elementary gate operations~\cite{Wu2013Geometric,single2, universal1, universal2} in several physical systems. However, the speed of the adiabatic quantum gates is rather slow, and thus decoherence will introduce considerable  errors~\cite{Wangxb2001Nonadiabatic,zhusl}.

To overcome the dilemma between the limited coherence times and the long duration of adiabatic evolution, implementation of quantum gates based on nonadiabatic geometric phases has been proposed ~\cite{Wangxb2001Nonadiabatic, zhusl, Sjoqvist2012Nonadiabatic, Xu2012Nonadiabatic}. Recently, in the non-Abelian case~\cite{Sjoqvist2012Nonadiabatic, Xu2012Nonadiabatic}, elementary quantum gates~\cite{Abdumalikov2013Experimental, nvsingle, long2017, ys2017, Zhou2017Holonomic, Xu2018Single,Yan2019Experimental, Feng2013Experimental,Zu2014Experimental,Nagata2018Universal, nmruniversal} have been experimentally demonstrated in various three-level physical systems. However, the noise-resilience of the geometric phases is not shared by this type of implementation~\cite{three1,three2,three3}. Indeed, robust quantum gates with non-Abelian geometric phases  can actually be implemented with two degenerated dark states~\cite{duan2001geometric, liubj}. However, it is experimentally difficult because of the need of complex control of quantum systems with four energy levels. On the other hand, experimental demonstration of universal quantum computation with nonadiabatic Abelian geometric phase is also lacking, due to the challenge of exquisite control among quantum systems. In addition, so far there is no direct experimental verification of the noise-resilient feature of  geometric quantum gates over the dynamical ones yet.

Here, with a multi-qubit superconducting quantum circuit architecture~\cite{clarke2008Superconducting, you2011Atomic, devoret2013superconducting}, we experimentally demonstrate a robust nonadiabatic geometric quantum computation~(GQC) scheme~\cite{GeoChenT, Zhao2017Rydberg}. The measured average fidelities for the realized single-qubit rotation gates and two-qubit controlled-Z~(CZ) gate are 0.9977(1) and 0.977(9), respectively, characterized by both quantum process tomography~(QPT) and randomized benchmarking~(RB) methods. The numbers in the brackets are the uncertainties obtained from repeated experiments of QPT and bootstrapping technique on the RB data, respectively. These gates are realized by merely using simple and experimentally accessible microwave controls over  capacitively-coupled superconducting transmon qubits, each of which involves only two states~\cite{koch2007chargeinsensitive}. The leakage of qubit states can be effectively suppressed and the coupling between the two qubits can be parametrically tuned in a large range~\cite{Lu2017Universal, Reagor2018Demonstration, TuneGExperiment, LiX2018Perfect}. Meanwhile, our demonstration only utilizes conventional resonant interaction for both single- and two-qubit gates, and thus simplifies the experimental complexity and decreases the error sources. Furthermore, we experimentally demonstrate the noise-resilient feature of the geometric quantum gates over the dynamical ones. Therefore, our experiment proves the way to achieve robust universal GQC on a large-scale qubit lattice.

We first explain how to construct the single-qubit geometric gate on a superconducting  qubit in the \{$\ket{0}$, $\ket{1}$\} subspace, where $\ket{0}$~($\ket{1}$) denotes the ground~(excited) state of the  qubit. Conventionally, single-qubit control is realized by applying a microwave drive on resonance with the qubit transition $\ket{0} \leftrightarrow \ket{1}$, as described by the Hamiltonian of
\begin{equation}
\label{SQDriveH}
H_1= \frac{1}{2}\Omega(t) e^{i\phi(t)}\ket{0}\bra{1} + H.c.,
\end{equation}
where $ \Omega(t) $ and $ \phi(t) $ are the time-dependent driving amplitude and phase of the microwave field. To achieve a universal set of  single-qubit nonadiabatic geometric gates in a single-loop way~\cite{Zhao2017Rydberg}, we divide the evolution time $\tau$ into three intervals: $0 \rightarrow \tau_1$, $\tau_1 \rightarrow \tau_2$, and $\tau_2 \rightarrow \tau$, with the driving amplitude and phase in each component satisfying
\begin{equation}
\label{GeoSQSplitPulse}
\begin{cases}
&\int_{0}^{\tau_1} \Omega(t)\mathrm{d}t=\theta,\quad \phi = \varphi-\frac{\pi}{2}, \quad t\in[0,\tau_1], \\
&\int_{\tau_1}^{\tau_2} \Omega(t)\mathrm{d}t=\pi,\quad \phi = \varphi+\gamma+\frac{\pi}{2}, \quad t\in[\tau_1,\tau_2],   \\
&\int_{\tau_2}^{\tau} \Omega(t)\mathrm{d}t=\pi - \theta,\quad \phi = \varphi-\frac{\pi}{2}, \quad t\in[\tau_2,\tau]. \\
\end{cases}
\end{equation}

\begin{figure}
    \includegraphics{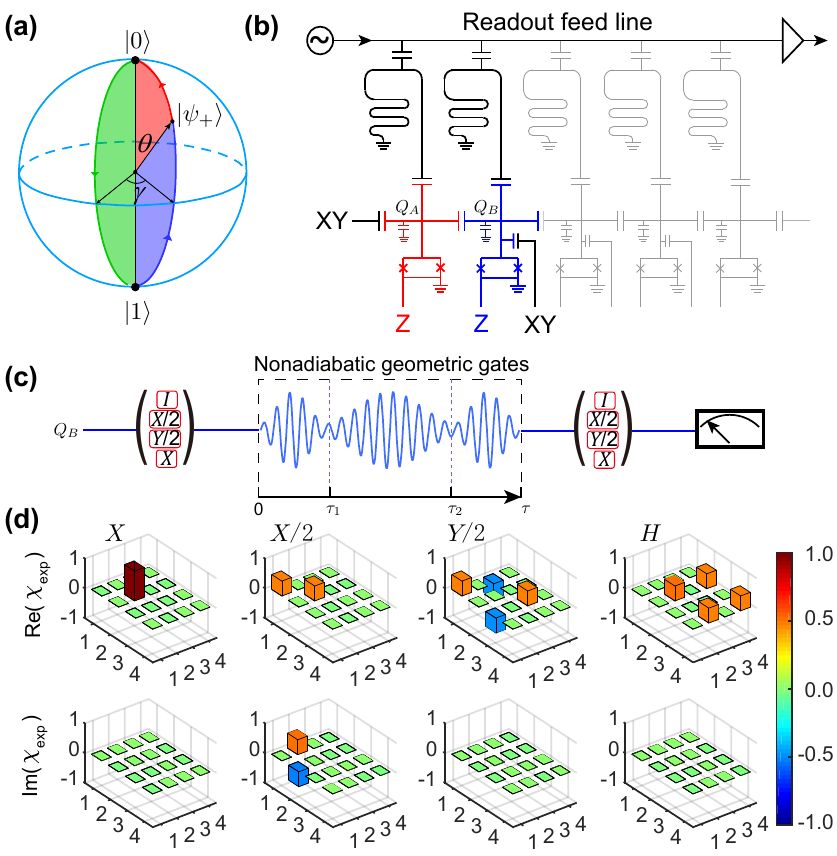}
    \caption{Single-qubit nonadiabatic geometric gates.  (a) Bloch sphere representation of the evolution trajectory to realize single-qubit geometric gates. (b) Simplified circuit schematic of the five-Xmon-qubit chain sample, with only the first two adjacent qubits $Q_A$ and $Q_B$ being considered in this work. (c) The experimental pulse sequence to characterize the performance of the single-qubit nonadiabatic geometric gates with the QPT method. The geometric gate is realized by three truncated Gaussian pulses with different amplitudes and phases. (d) Bar charts of the real and imaginary parts of $\chi_\mathrm{exp}$ of four specific gates: $X$, $X/2$, $Y/2$, and Hadamard $H$, giving an average process fidelity of 0.9980(14). The numbers in the $x$ and $y$ axes correspond to the operators in the basis set \{$I$, $\sigma_x$, $-i\sigma_y$, $\sigma_z$\} in the \{\ket{0}, \ket{1}\} subspace. The solid black outlines are for the ideal gates.}
    \label{fig:fig1}
\end{figure}

Consequently, two orthogonal states $\ket{\psi_+} = \cos \frac{\theta}{2} \ket{0} + \sin \frac{\theta}{2} e^{i \varphi} \ket{1}$ and $\ket{\psi_-} = \sin \frac{\theta}{2} e^{-i \varphi} \ket{0} - \cos \frac{\theta}{2}  \ket{1}$ undergo a cyclic orange-slice-shaped evolution on the single-qubit Bloch sphere~\cite{Tian2004Geometric}, as shown in Fig.~\ref{fig:fig1}(a), resulting in a geometric phase $\gamma$~($-\gamma$) on the quantum state $\ket{\psi_+}$~($\ket{\psi_-}$). We note that this construction can be recognized as a special type of composite pulses, but whose robustness is originated from the pure geometric nature~\cite{Ota2009,Ichikawa2012}. This construction is however different from other traditional composite pulses~\cite{Levitt1986, Vandersypen2005, Jones2011}, where complex concatenated pulses are optimized to compensate the specific error for a certain gate and a larger pulse area than our scheme is generally required, resulting in a higher gate infidelity from decoherence. Therefore, the obtained single-qubit gate of the total geometric evolution is
\begin{eqnarray} \label{GeoSQU}
U_1 \left( \theta, \gamma, \varphi \right) &=& \cos\gamma + i\sin\gamma \left(
\begin{array}{cc}
\cos\theta  &   \sin\theta e^{-i\varphi}    \\
\sin\theta e^{i\varphi} &   -\cos\theta
\end{array}
\right)\notag\\
&=& \exp{\left( i\gamma \vec{n} \cdot \vec{\sigma} \right)},
\end{eqnarray}
which corresponds to a rotation operation around the axis $ \vec{n}=(\sin\theta\cos\varphi,\sin\theta\sin\varphi,\cos\theta) $  by an angle $ -2\gamma $. The parameters $ \theta $, $ \gamma $, $ \varphi $ are determined by the drive.

Our experiment is performed on a five-Xmon-qubit chain sample~\cite{LiX2018Perfect,CaiW2019Observation}, with the simplified circuit schematic shown in Fig.~\ref{fig:fig1}(b). Only two adjacent qubits $Q_A$ and $Q_B$ are used in this experiment, with $ \ket{0} \leftrightarrow \ket{1} $ transition frequency of $ \omega_A/2\pi = 4.602 $ GHz and $ \omega_B/2\pi = 5.081 $ GHz, respectively, and a static capacitive coupling strength $g_{AB}/2\pi \approx 17$~MHz between them. Only the lowest two energy levels are considered here due to the large anharmonicity $\alpha_A/2\pi = -202$~MHz and $\alpha_B/2\pi = -190$~MHz for the Xmon qubits $Q_A$ and $Q_B$, respectively. Each qubit has individual XY and Z drive lines for qubit state manipulation and frequency tunability, and is coupled to a separate $\lambda/4$ resonator for individual and simultaneous readout. 
More details about the experimental setup and device parameters can be found in Ref.~\cite{Supplement}.

We first demonstrate the single-qubit nonadiabatic geometric gates on  qubit $Q_B$, with the experimental pulse sequence shown in Fig.~\ref{fig:fig1}(c). As a demonstration, here we fix $\theta = \pi / 2$, and realize single-qubit $\pi$ and $\pi/2$ rotations around $X$ and $Y$ axes~(denoted as $X$, $Y$, $X/2$, and $Y/2$ respectively), which construct a basis set to generate single-qubit Cliffords. The geometric gate consists of a $\pi$ rotation sandwiched by two $\pi /2 $ rotations with a total width of 80~ns. The envelope of each pulse is a truncated Gaussian pulse with the correction of ``derivative removal by adiabatic gate" method in order to suppress the leakage to the undesired energy levels~\cite{Motzoi}.

We first characterize the single-qubit geometric gates by the QPT method~\cite{Supplement}, with the experimental sequence shown in Fig.~\ref{fig:fig1}(c). The experimental process matrices $\chi_\mathrm{exp}$ of four specific geometric gates $X$, $X/2$, $Y/2$ and Hadamard $H$~(implemented with a $Y/2$ rotation followed by a $X$ rotation) are shown in Fig.~\ref{fig:fig1}(d) with an average process fidelity of 0.9980(14). The process fidelity is calculated through $F_p = \mathrm{Tr} \left( \chi_\mathrm{exp} \chi_\mathrm{ideal} \right)$, where $\chi_\mathrm{ideal}$ is the ideal process matrix for the corresponding gate.

\begin{figure}
    \includegraphics{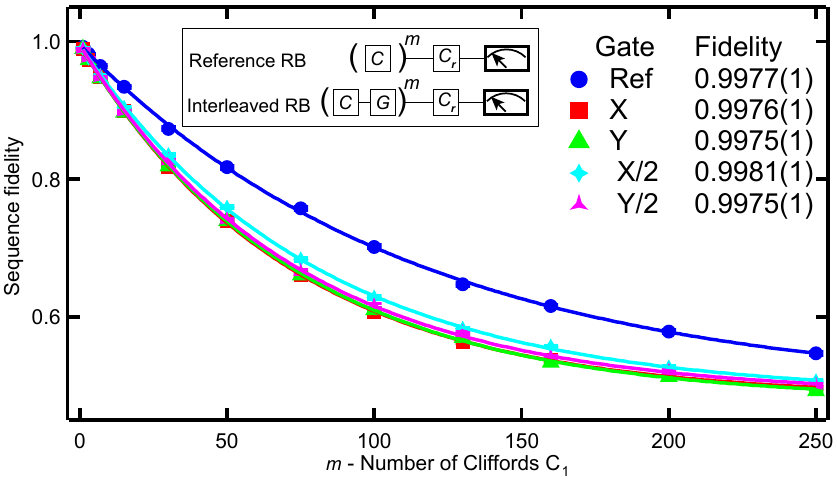}
    \caption{RB of single-qubit nonadiabatic geometric gates. Inset is the experimental pulse sequences to perform both the reference RB and interleaved RB experiments. Fit to the reference decay curve gives an average fidelity of 0.9977(1) for the single-qubit geometric gates in the Clifford group. The difference between the reference and the interleaved decay curves gives the gate fidelity of four specific gates: $X$, $Y$, $X/2$, $Y/2$.}
    \label{fig:fig2}
\end{figure}

Another conventional method, Clifford-based RB~\cite{RBSQProtocol, RBMultiQProtocol, RBInterleaved}, is also used to characterize the geometric gates, with the sequences for both the reference RB and interleaved RB experiments shown in the inset of Fig.~\ref{fig:fig2}. The experimentally measured ground state probability~(the sequence fidelity) decays as a function of the number of single-qubit Cliffords $m$ for both the reference RB and interleaved RB experiments are shown in Fig.~\ref{fig:fig2}. Both curves are fitted to $F = Ap^m +B$ with different sequence decays $p=p_\mathrm{ref}$ and $p=p_\mathrm{gate}$. The reference RB experiment gives an average fidelity $F_\mathrm{avg}=1 - (1 - p_\mathrm{ref})/3.75=0.9977(1)$ for the realized single-qubit nonadiabatic geometric gates in the Clifford group. The measured interleaved gate fidelities $F_\mathrm{gate} = 1 - (1 - p_\mathrm{gate}/p_\mathrm{ref})/2$ of the four specific  geometric gates $X$, $Y$, $X/2$, and $Y/2$, inserted in the random Cliffords in the interleaved RB experiment, are 0.9976(1), 0.9975(1), 0.9981(1), and 0.9975(1), respectively.

\begin{figure}
    \includegraphics{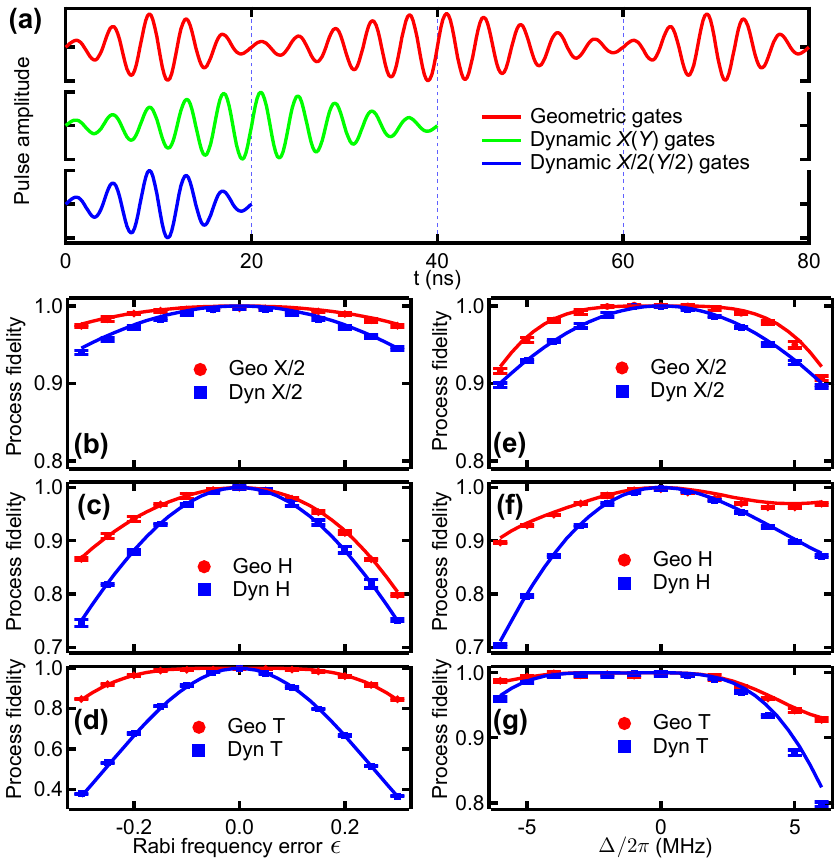}
    \caption{Noise-resilient feature of single-qubit geometric gates. (a) Pulse shapes of both geometric gates and dynamical gates, which are constrained to have the same driving strength. Hadamard gate Geo $H$~(Dyn $H$) is implemented with a geometric~(dynamical) $Y/2$ rotation followed by a geometric~(dynamical) $X$ rotation, while $T$ phase gate Geo $T$~(Dyn $T$) is realized with a geometric~(dynamical) $X$ rotation followed by a geometric~(dynamical) $\pi$ pulse along an axis in the $xy$ plane with an angle of $\pi/8$ to the $x$ axis. (b-d) The experimental process fidelities of single-qubit gates: $X/2$~(b), $H$~(c), and $T$~(d) realized by both geometric and dynamical means, as a function of Rabi frequency error $\epsilon$. The experimental results are also consistent with the numerical simulations~(solid lines). (e-g) The experimental process fidelities of single-qubit gates: $X/2$~(e), $H$~(f), and $T$~(g) realized by both geometric and dynamical means, as a function of qubit frequency detuning $\Delta$, consistent with the numerical simulation results~(solid lines).}
    \label{fig:fig3}
\end{figure}

With the realized single-qubit nonadiabatic geometric gates, we further demonstrate their robustness against two different types of errors: control amplitude error and qubit frequency shift-induced error, which will be the dominant gate error sources for a large scale qubit lattice. In our experiment, we compare the geometric gates with the conventional dynamical gates under the same driving strength, with the pulse envelopes shown in Fig.~\ref{fig:fig3}(a). We have experimentally characterized the performance of three geometric gates: $X/2$, $H$, and $T$ phase gate with a single-qubit QPT method, as a function of Rabi frequency error $\epsilon$~(a relative offset in Rabi frequency) and qubit frequency detuning $\Delta$, as well as that for the corresponding dynamical gates. The experimentally measured process fidelities as a function of these two errors are shown in Figs.~\ref{fig:fig3}(b-g).

The geometric gates are realized with two different configuration settings, corresponding to two different geometric evolution trajectories. In configuration A, the geometric gates are realized with the geometric evolution described in Eq.~\ref{GeoSQSplitPulse} and have distinct advantages over the dynamical gates against additional Rabi frequency error $\epsilon$, as shown in Figs.~\ref{fig:fig3}(b-d). In configuration B, the geometric gates are realized by setting the phase $\phi = \varphi + \gamma -\pi/2$ at [$\tau_1$, $\tau_2$] interval in Eq.~\ref{GeoSQSplitPulse}, while the unitary of the geometric gate remains the same as that in Eq.~\ref{GeoSQU} when $\theta = \pi/2$. The noise-resilient feature of the geometric gates still persists for different detuning errors, as shown in Figs.~\ref{fig:fig3}(e-g). All experimental results also agree very well with the numerical simulations. The comparisons clearly illustrate the distinct advantages of the realized nonadiabatic geometric gates. We note that the noise-resilient feature of the geometric gates depends on the types of errors and the cyclic evolution paths of the geometric gates~\cite{Supplement}. The geometric gates realized with configuration A do not always outperform the dynamical gates with additional frequency detuning errors, and the geometric gates realized with configuration B also do not perform better than the dynamical gates with different Rabi frequency errors. However, one can always find a specific evolution path of the control pulse to realize a noise-resilient geometric gate against the dominant error in the system.

In order to achieve a universal quantum computation, two-qubit entangling operations are also necessary. In our experiment, a non-trivial two-qubit geometric gate is also realized in a similar way to the single-qubit case by using a parametric modulation drive of one qubit frequency. Considering two adjacent qubits $Q_A$ and $Q_B$~(with anharmonicities $\alpha_A$ and $\alpha_B$) capacitively coupled to each other, the qubit frequency of $Q_A$ is modulated with a sinusoidal form:
$\omega_A(t)=\omega_A + \varepsilon \sin(\nu t + \Phi)$,
where $\omega_A$ is the mean operating frequency, and $\varepsilon$, $\nu$, and $\Phi$ are the modulation amplitude, frequency, and phase, respectively. Ignoring the higher-order oscillating terms, when the modulation frequency satisfies $\nu = \omega_B - \omega_A + \alpha_B $, the parametric drive will induce a transition operation between the two energy levels $\ket{11} \leftrightarrow \ket{02}$ in the two-qubit subspace with the effective Hamiltonian in the interaction picture  as
\begin{equation}
\label{TuneG1102Hamiltonian}
H_2=\frac{1}{2} \tilde{g} e^{i\tilde{\phi}} \ket{11}\bra{02}+H.c.,
\end{equation}
where $\tilde{g} = 2g_{AB}J_1\left( {\varepsilon}/{\nu} \right)$ and $\tilde{\phi} = -\Phi + \pi/2$ are the effective coupling strength and phase of the parametric drive, with $J_1\left( {\varepsilon}/{\nu} \right)$ being the $1^\mathrm{st}$ order Bessel function of the first kind. Similar to the single-qubit geometric gates with the Hamiltonian of Eq.~\ref{SQDriveH}, we can realize arbitrary geometric gates in the subspace \{$\ket{11}$, $\ket{02}$\} by modulating the effective coupling strength and phase in three time intervals. As a demonstration, we fix $\theta = 0$, resulting in two time intervals of the gate, and realize the geometric phase gate
$\left( \begin{array}{cc}
e^{i\gamma} &   0   \\
0   &   e^{-i\gamma}
\end{array} \right)$ in the subspace. When only considering the unitary in the two-qubit computational space \{$\ket{00}$, $\ket{01}$, $\ket{10}$, $\ket{11}$\}, the resulting unitary operation corresponds to a controlled-phase gate with an entangled phase $\gamma$:
\begin{equation}
\label{GeoCZU01}
U_2(\gamma) = \left(
\begin{array}{cccc}
1   &   0   &   0   &   0   \\
0   &   1   &   0   &   0   \\
0   &   0   &   1   &   0   \\
0   &   0   &   0   &   e^{i\gamma}
\end{array}
\right).
\end{equation}

\begin{figure}
    \includegraphics{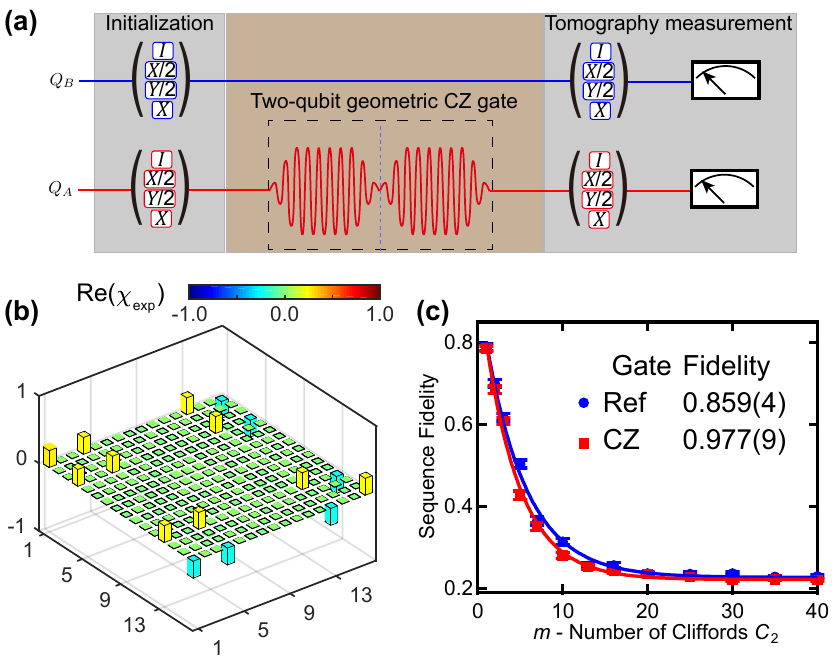}
    \caption{Two-qubit geometric CZ gate. (a) Experimental pulse sequence to perform two-qubit QPT of the geometric CZ gate, which is realized with two square pulses with additional rising and falling edges~(black dotted box). (b) Real part of the experimental process matrix$ \chi_\mathrm{exp} $ for the geometric CZ gate, giving a process fidelity of $0.941(13)$. Measured imaginary part is smaller than 0.09 and not shown. The numbers in the $x$ and $y$ axes correspond to the operators in the basis set \{$I$, $\sigma_x$, $-i\sigma_y$, $\sigma_z$\}$^{\otimes 2}$ in the \{\ket{00}, \ket{01}, \ket{10}, \ket{11}\} subspace. The solid black outlines are for the ideal CZ gate. (c) Two-qubit RB data of the geometric CZ gate between qubits $Q_A$ and $Q_B$, with an extracted $F_\mathrm{CZ} = 0.977(9)$.}
    \label{fig:fig4}
\end{figure}

The two-qubit geometric controlled-phase gate is performed with two sinusoidal modulation drives applied in series. Each has a square pulse envelope with sine squared rising and falling edges to suppress the adverse impact of sudden phase changes. The modulation frequency $\nu/2\pi = 268.2$~MHz and the modulation amplitude $\varepsilon/2\pi = 150$~MHz lead to an effective coupling strength $\tilde{g}/2\pi \approx 10$~MHz. Thus, the two-qubit gate is implemented with a duration of 112.8~ns. As an example, we here fix $\gamma = \pi$ and realize a CZ gate for the two qubits. We first use the two-qubit QPT method to benchmark the performance of the realized CZ gate, with the experimental sequence shown in Fig.~\ref{fig:fig4}(a). The experimentally reconstructed process matrix $\chi_\mathrm{exp}$ is shown in Fig.~\ref{fig:fig4}(b) and indicates a process fidelity of 0.941(13) for the realized geometric CZ gate.

Besides, a two-qubit Clifford-based RB experiment is also performed to characterize the fidelity of the realized geometric CZ gate. The final measured ground state probability~(sequence fidelity) decays as a function of the number of two-qubit Cliffords are displayed in Fig.~\ref{fig:fig4}(c) for both the two-qubit reference RB and CZ-interleaved RB experiments. We extract the geometric CZ gate fidelity $F_\mathrm{CZ} = 1 - \frac{3}{4}\left( 1 - \frac{p_\mathrm{CZ}}{p_\mathrm{ref}} \right) = 0.977(9)$ from fitting both the reference and interleaved RB decay curves. This result is consistent with that from the two-qubit QPT method, when considering the state preparation and measurement error of about 0.03. The infidelity of the CZ gate mainly comes from the decoherence of the two qubits, also confirmed with our numerical simulations. The extracted average Clifford fidelity $F_\mathrm{C_2}$ = 0.859(4), mainly limited by qubit decoherence and crosstalk between the two qubits~\cite{Supplement}.

In conclusion, we experimentally realize single-qubit nonadiabatic geometric gates with an average fidelity of 0.9977(1). The noise-resilient feature of the realized single-qubit geometric gates is also verified by comparing the performances of both the  geometric and dynamical gates with different errors. In addition,  a two-qubit nonadiabatic geometric CZ gate is also implemented with a fidelity of 0.977(9). Therefore, the demonstrated universal geometric quantum gate set opens the door to implement high-fidelity quantum gates for robust geometric quantum computation.

\begin{acknowledgments}
Z.Y.X. and T.C. are supported in part by the Key R\&D Program of Guangdong Province (Grant No. 2018B030326001), the National Natural Science Foundation of China (Grant No. 11874156), and the National Key R\&D Program of China (Grant No. 2016 YFA0301803). L.S. is supported by National Key Research and Development Program of China No.2017YFA0304303 and the National Natural Science Foundation of China under Grants No.11874235.
\end{acknowledgments}

\textit{Note added.}-- While we were preparing our manuscript, we noticed a similar implementation of nonadiabatic single-qubit geometric gates with a superconducting qubit~\cite{zhao2019experimental}.


%

\end{document}


\title{Supplementary Material for ``Experimental implementation of universal nonadiabatic geometric quantum gates in a superconducting circuit"}

\author{Y. Xu}
\thanks{These three authors contributed equally to this work.}
\affiliation{Center for Quantum Information, Institute for Interdisciplinary Information
Sciences, Tsinghua University, Beijing 100084, China}
\author{Z. Hua}
\thanks{These three authors contributed equally to this work.}
\affiliation{Center for Quantum Information, Institute for Interdisciplinary Information
Sciences, Tsinghua University, Beijing 100084, China}
\author{Tao Chen}
\thanks{These three authors contributed equally to this work.}
\affiliation{Guangdong Provincial Key Laboratory of Quantum Engineering and Quantum Materials, GPETR Center for Quantum Precision Measurement, and School of Physics and Telecommunication Engineering, South China Normal University, Guangzhou 510006, China}
\author{X. Pan}
\affiliation{Center for Quantum Information, Institute for Interdisciplinary Information
Sciences, Tsinghua University, Beijing 100084, China}
\author{X. Li}
\affiliation{Center for Quantum Information, Institute for Interdisciplinary Information
Sciences, Tsinghua University, Beijing 100084, China}

\author{J. Han}
\author{W. Cai}
\author{Y. Ma}
\author{H. Wang}
\author{Y.P. Song}
\affiliation{Center for Quantum Information, Institute for Interdisciplinary Information Sciences, Tsinghua University, Beijing 100084, China}

\author{Zheng-Yuan Xue} \email{zyxue83@163.com}
\affiliation{Guangdong Provincial Key Laboratory of Quantum Engineering and Quantum Materials, GPETR Center for Quantum Precision Measurement, and School of Physics and Telecommunication Engineering, South China Normal University, Guangzhou 510006, China}

\author{L.~Sun} \email{luyansun@tsinghua.edu.cn}
\affiliation{Center for Quantum Information, Institute for Interdisciplinary Information
Sciences, Tsinghua University, Beijing 100084, China}

\pacs{} \maketitle

\section{Experimental Setup}
The sample in our experiment is a five-qubit chain device, which consists of five adjacent cross-shaped transmon (Xmon) qubits~\cite{Xmon2013}, arranged in a linear array with nearly identical nearest-neighbor coupling strengths. We place the experimental device inside a dilution refrigerator with a base temperature of about 10 mK. The detail of the device has been described in Refs~\cite{LiX2018Perfect,CaiW2019Observation}. 

In the experiment, we have performed the geometric gates with only the first two capacitively coupled qubits $Q_A$ and $Q_B$, whose main parameters are summarized and listed in Table~\ref{Table:parameters}. The other three qubits are biased far away from these two operation qubits and thus are nearly completely decoupled. 

\begin{table}
	\caption{Device parameters of the two operating qubits.}
	\begin{tabular}{cp{0.3cm}<{\centering}p{0.3cm}<{\centering}p{0.6cm}<{\centering}p{0.6cm}<{\centering}}
		&&&& \tabularnewline
		\hline
		\hline
		&&&&\\[-0.9em]
		Parameters &\multicolumn{2}{c}{$Q_A$}  &\multicolumn{2}{c}{$Q_B$}  \tabularnewline
		\hline
		&&&&\\[-0.9em]
		Readout frequency (GHz) &\multicolumn{2}{c}{$6.8386$} &\multicolumn{2}{c}{$6.8633$}   \tabularnewline
		Qubit frequency (GHz) (sweet spot) &\multicolumn{2}{c}{$4.7813$} &\multicolumn{2}{c}{$5.1261$}  \tabularnewline
		Qubit frequency (GHz) (operating spot) &\multicolumn{2}{c}{$4.6019$} &\multicolumn{2}{c}{$5.0810$}  \tabularnewline
		Anharmonicity~($\alpha_A/2\pi$, $\alpha_B/2\pi$) (MHz) &\multicolumn{2}{c}{-202} &\multicolumn{2}{c}{-190} \tabularnewline
		$T_1$ ($\mu$s) (sweet spot) &\multicolumn{2}{c}{$15.4$} &\multicolumn{2}{c}{$11.2$} \tabularnewline
		$T^*_2$ ($\mu$s) (sweet spot) &\multicolumn{2}{c}{$9.86$} &\multicolumn{2}{c}{$16.3$} \tabularnewline
		$T_{\mathrm{2E}}$ ($\mu$s) (sweet spot) &\multicolumn{2}{c}{$17.3$} &\multicolumn{2}{c}{$20.9$} \tabularnewline
		$T_1$ ($\mu$s) (operating point) &\multicolumn{2}{c}{$20.5$} &\multicolumn{2}{c}{$26.1$} \tabularnewline
		$T^*_2$ ($\mu$s) (operating point) &\multicolumn{2}{c}{$1.73$} &\multicolumn{2}{c}{$4.86$} \tabularnewline
		Qubit-qubit coupling strength ${\rm g}_{AB}/2\pi$ (MHz) & &\multicolumn{2}{c}{$16.68$} & \tabularnewline
		Qubit-readout dispersive shift $\rm\chi_{qr}/2\pi$ (MHz) &\multicolumn{2}{c}{$0.09$} &\multicolumn{2}{c}{$0.12$} \tabularnewline
		Readout resonator decay rate $\rm\kappa_r/2\pi$ (MHz) &\multicolumn{2}{c}{$0.88$} &\multicolumn{2}{c}{$1.06$} \tabularnewline
		\hline
		\hline
	\end{tabular} \vspace{-6pt}
	\label{Table:parameters}
\end{table}

\begin{figure*}[tbh]
	\includegraphics[width=0.75\linewidth]{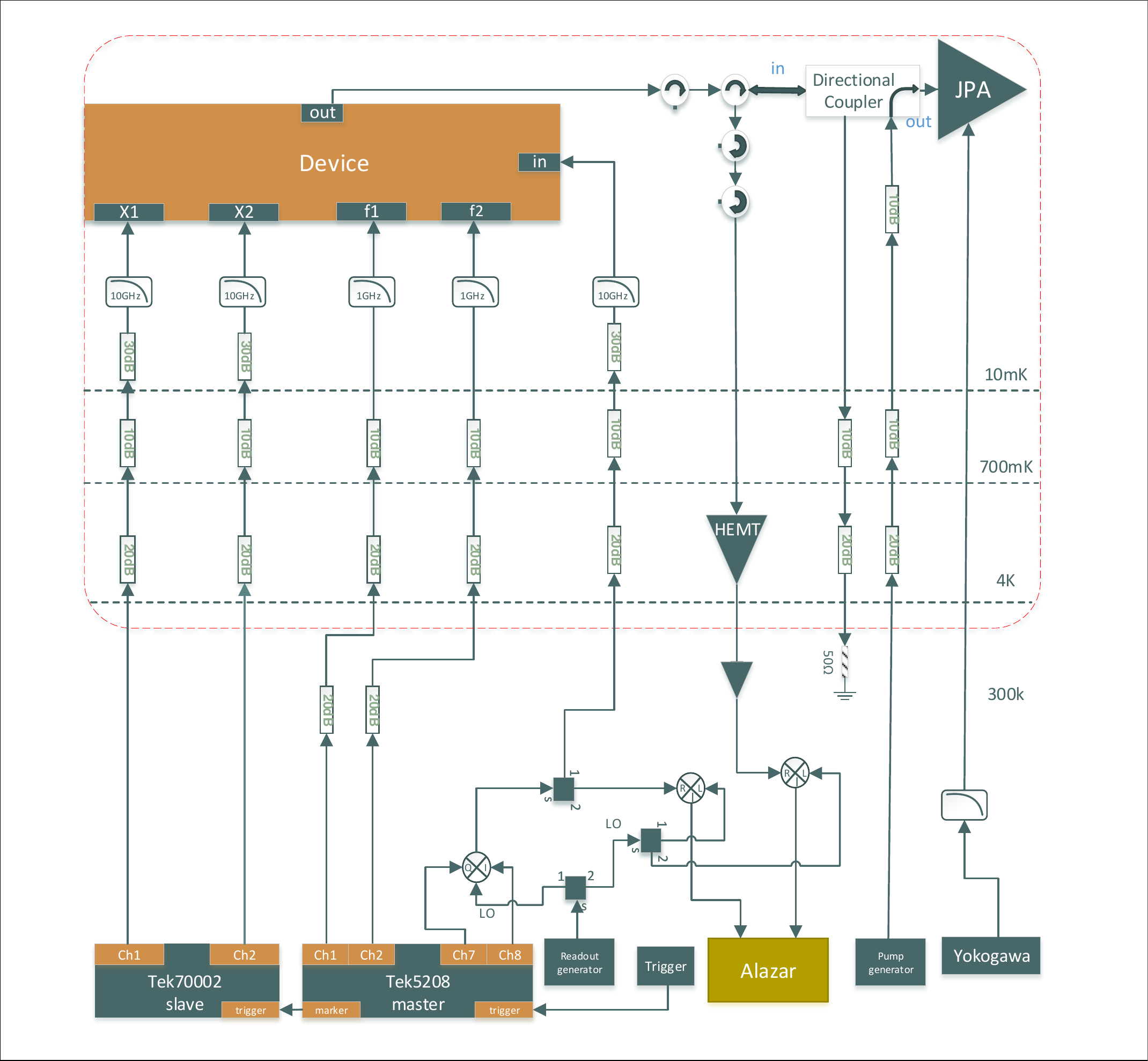}
	\caption{Details of wiring and circuit components.}
	\label{fig:diagramgeogate1905}	
\end{figure*}

Details of the measurement circuitry of our experiment is shown in Fig.~\ref{fig:diagramgeogate1905}. The XY rotations of the two qubits are controlled by microwave pulses directly generated from a two-channel arbitrary waveform generator AWG70002A without extra IQ modulation, benefiting from its large bandwidth and sampling rate. The Z controls of the two qubits are performed with the first two channels of another synchronized eight-channel AWG5208. The last two channels of AWG5208 are used to provide a pair of sideband modulations at different frequencies for the readout of the two qubits, in combination with a signal generator as the local oscillator (LO). The transmitted readout signal is first amplified by a Josephson parametric amplifier (JPA) at the base temperature of the dilution refrigerator. The JPA with an amplification gain of 20~dB and a bandwidth about 260~MHz allows high-fidelity single-shot measurements of the two qubits individually and simultaneously. A high-electron-mobility-transistor (HEMT) amplifier at 4 K and a standard commercial low-noise microwave amplifier at room temperature are also used before the down-conversion of the readout signal to the applied sideband frequencies with the same readout generator as the LO. After these amplications and down-conversion, the readout signal is finally digitized and recorded by an Alazar card, as well as the reference signal without going through the dilution refrigerator. 

Crosstalk between Z control lines of the two qubits is inevitable due to the ground plane return currents. We compensate these flux crosstalks by orthogonalization of the flux bias lines through an inversion of the normalized qubit frequency response matrix
\begin{equation}
\label{equ:FluxMatrix}
M = \left(  
\begin{array}{cc}
1	&	-0.0759		\\
0.0800	&	1	
\end{array}	
\right).
\end{equation}
Thus by performing the inversion of $M$ to the flux bias controls, each qubit frequency can be independently controlled without changing the other qubit frequency.

\begin{table}[b]
	\caption{Two-qubit simultaneous readout assignment probability matrix $\mathcal{R}$. Each column represents the two-qubit measurement probabilities after preparing the qubits in the corresponding basis state. }
	\begin{tabular}{cp{1.5cm}<{\centering}p{1.5cm}<{\centering}p{1.5cm}<{\centering}p{1.5cm}<{\centering}p{1.5cm}<{\centering}}
		&&&& \tabularnewline
		\hline
		\hline
		&&&&\\[-0.9em]
		        &  $\ket{00}$     &  $\ket{01}$ & $\ket{10}$ & $\ket{11}$ \tabularnewline
		\hline
		&&&&\\[-0.9em]
		00   &  0.9918  &  0.1058  &  0.1279  &  0.0131  \tabularnewline
		01   &  0.0031  &  0.8905  &  0.0005  &  0.1137  \tabularnewline
		10   &  0.0051  &  0.0006  &  0.8686  &  0.0890  \tabularnewline
		11   &  0.0000  &  0.0032  &  0.0030  &  0.7842  \tabularnewline
		\hline
		\hline
	\end{tabular} \vspace{-6pt}
	\label{Table:readout}
\end{table}

With the quantum-limited JPA, both qubits can be readout individually and simultaneously with high fidelity in single-shot measurements. Due to the qubit thermal excitation and relaxation during measurements, there are non-negilable readout infidelities. In order to calibrate the two-qubit readout error, we prepare the system in each computational basis state and simultaneously measure the assignment probability $\vec{p} = \left(p_{00}, p_{01}, p_{10}, p_{11}\right)^T$ of the two qubits. By repeating the experiments for all the two-qubit computational basis states, we obtain the $4 \times 4$ readout matrix $\mathcal{R}$ as shown in Table~\ref{Table:readout}. Thus the readout errors can be corrected by multiplying the inverse of the readout matrix $\mathcal{R}$ with the measured probability $\vec{p}$, such that $\vec{p}_\mathrm{corr} = \mathcal{R}^{-1}\cdot \vec{p}$ represents the real occupation probabilities of the four computational basis states.

\section{Quantum process tomography}
Both single- and two-qubit geometric gates are firstly characterized with quantum process tomography~(QPT) method in the main text.

In the single-qubit QPT experiment with the experimental sequence shown in Fig.~1(c) of the main text, we first initialize the qubit with the following four states \{ $\ket{0}$, $\ket{1}$, $(\ket{0} + \ket{1})/\sqrt{2}$, $(\ket{0} - i \ket{1})/\sqrt{2}$ \}, then apply the nonadiabatic geometric gates, and finally perform the state tomography measurements of the final states. The density matrix of the final state is reconstructed by four pre-rotations \{$I$, $X/2$, $Y/2$, $X$\} in the state tomography measurement, where $I$ represents the identity operator. With the four initial states $\rho_i$, the experimental process matrix $\chi_\mathrm{exp}$ can be extracted from the four corresponding final states $\rho_f$ through $\rho_f = \sum_{m,n}{\chi_{mn}E_m\rho_iE_n^{\dag}}$~\cite{Nielsen}, where the basis operators $E_m$ and $E_n$ are chosen from \{ $I$, $\sigma_x$, $-i\sigma_y$, $\sigma_z$ \} with $\sigma_x$, $\sigma_y$, and $\sigma_z$ being Pauli operators.

The two-qubit QPT experiment is similar to that for the single-qubit case but with 16 initial states chosen from \{$\ket{0}$, $\ket{1}$, $\left( \ket{0} + \ket{1} \right)/\sqrt{2}$, $\left( \ket{0} -i \ket{1} \right)/\sqrt{2}$\}$^{\otimes 2}$ instead. The density matrix of the final two-qubit state is reconstructed from the state tomography measurements with 16 pre-rotations \{ $I$, $X/2$, $Y/2$, $X$\}$^{\otimes 2}$.

\section{Randomized benchmarking}
We also use another conventional method, Clifford-based randomized benchmarking~(RB), to characterize the geometric single-qubit rotation gates and two-qubit CZ gate.

In the single-qubit RB experiment, we perform both the reference RB and interleaved RB experiments with the experimental sequences shown in the inset of Fig.~2 of the main text. In the reference RB experiment, we first apply a random sequence of $m$ quantum gates chosen from the single-qubit Clifford group~($C_1$), then append a recovery gate $C_r$ to invert the whole sequence, and finally measure the ground state probability as the sequence fidelity. The whole experiment is repeated for $k = 50$ different sequences to get the average sequence fidelity. In the interleaved RB experiment, a specific gate $G$ is interleaved into the $m$ random Cliffords, and a similar recovery gate is applied to invert the whole sequence. The experimentally measured sequence fidelity curves as a function of the number of Cliffords $m$ for both reference RB and interleaved RB experiments are fitted to $F = Ap^m +B$ with different sequence decays $p=p_\mathrm{ref}$ and $p=p_\mathrm{gate}$. The average gate fidelity is given by $F_\mathrm{ref} = 1 - (1 - p_\mathrm{ref})(d-1)/d/1.875$, with $d = 2^N$ for $N$ qubits. The difference between the reference and interleaved RB experiments gives the specific gate fidelity $F_\mathrm{gate} = 1 - (1 - p_\mathrm{gate}/p_\mathrm{ref})(d-1)/d$.
 
The two-qubit Clifford-based RB experiment is similar, but with the random gates chosen from the two-qubit Clifford group~($C_2$) instead. For the CZ-interleaved RB experiment, the geometric CZ gate is inserted into the random sequence.

\section{Comparison between geometric and dynamical gates}
We have demonstrated the robustness of the single-qubit geometric gates against two different types of errors (i.e. control amplitude error and qubit frequency shift-induced error) by comparing the geometric gates and the conventional dynamical gates with the same driving strength. The geometric gates are realized with two different configuration settings, corresponding to two different geometric evolution trajectories. 

In configuration A, the geometric gates are realized with a three-component microwave drive to generate a cyclic geometric evolution. The driving strengths and phases of the three time intervals are described as
\begin{equation}
\label{GeoSQSplitPulse1}
\begin{cases}
&\int_{0}^{\tau_1} \Omega(t)\mathrm{d}t=\theta,\quad \phi = \varphi-\frac{\pi}{2}, \quad t\in[0,\tau_1],	\\
&\int_{\tau_1}^{\tau_2} \Omega(t)\mathrm{d}t=\pi,\quad \phi = \varphi+\gamma+\frac{\pi}{2}, \quad t\in[\tau_1,\tau_2],	\\
&\int_{\tau_2}^{\tau} \Omega(t)\mathrm{d}t=\pi - \theta,\quad \phi = \varphi-\frac{\pi}{2}, \quad t\in[\tau_2,\tau]. \\
\end{cases}
\end{equation}

\begin{figure}
	\includegraphics{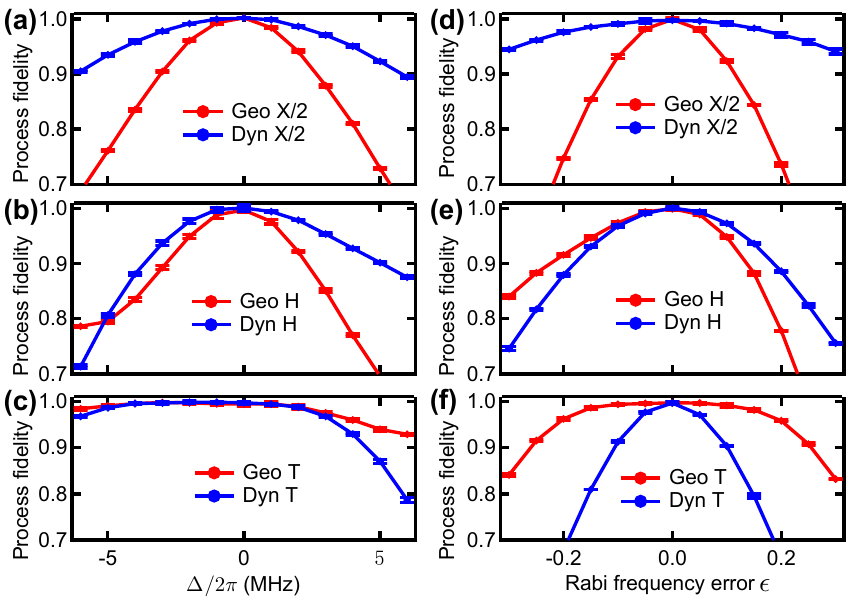}
	\caption{Comparison between geometric and dynamical gates. (a-c) The experimental process fidelity of single-qubit gates: $X/2$~(a), $H$~(b), and $T$~(c), realized by both geometric and dynamical means, as a function of qubit frequency detuning. The geometric gates, realized with configuration A, do not always perform better than the dynamical gates with detuning errors. (d-f) The experimental process fidelity of single-qubit gates: $X/2$~(d), $H$~(e), and $T$~(f), realized by both geometric and dynamical means, as a function of Rabi frequency error $\epsilon$. The geometric gates, realized with configuration B, do not always perform better than the dynamical gates with Rabi frequency errors.}
	\label{fig:SQPT_Geo_Dyn_A_hx_H_T}
\end{figure}

We have shown the realized geometric gates in configuration A exhibit distinct advantages over the conventional dynamical gates with additional Rabi frequency error $\epsilon$, as shown in the main text. However, we note that the geometric gates in configuration A do not always outperform the dynamical gates when introducing additional frequency detuning errors, as shown in Figs.~\ref{fig:SQPT_Geo_Dyn_A_hx_H_T}(a-c). 

In configuration B, the geometric gates are also realized with a three-component microwave drive, but with different phase in the second interval, described as
\begin{equation}
\label{GeoSQSplitPulse2}
\begin{cases}
&\int_{0}^{\tau_1} \Omega(t)\mathrm{d}t=\theta,\quad \phi = \varphi-\frac{\pi}{2}, \quad t\in[0,\tau_1],	\\
&\int_{\tau_1}^{\tau_2} \Omega(t)\mathrm{d}t=\pi,\quad \phi = \varphi+\gamma-\frac{\pi}{2}, \quad t\in[\tau_1,\tau_2],	\\
&\int_{\tau_2}^{\tau} \Omega(t)\mathrm{d}t=\pi - \theta,\quad \phi = \varphi-\frac{\pi}{2}, \quad t\in[\tau_2,\tau]. \\
\end{cases}
\end{equation}

The unitary gate of this geometric path is the same as that of configuration A when $\theta = \pi/2$. We have demonstrated the noise-resilient feature of this gate with different frequency detuning errors in the main text. However, the geometric gates in this configuration do not always perform better than the dynamical gates with different Rabi frequency errors as well, as shown in Figs.~\ref{fig:SQPT_Geo_Dyn_A_hx_H_T}(d-f).  

Therefore, we conclude that the noise-resilient feature of the geometric gates depends on the types of errors and the cyclic evolution paths. Most time one may not find appropriate parameters to realize geometric gates against all kinds of errors, but one can still find a specific evolution path to realize noise-resilient geometric gates against the dominant error in the system.  

\section{Calibration of Geometric CZ}
In order to achieve high-fidelity gemetric CZ gate, a few experimental parameters should be carefully calibrated. In this section, we will describe in detail the calibration procedure of the realized two-qubit geometric CZ gate.

In our experiment, the two-qubit CZ gate is realized with a parametric modulation drive of one qubit frequency. After initializing the two qubits in $\ket{11}$ state, we modulate the qubit frequency of $Q_A$ with a sinusoidal form of 
\begin{equation}
\label{TuneSinFrequency}
\omega_A(t)=\omega_A + \varepsilon \sin(\nu t + \Phi),
\end{equation}
for different time durations, and finally measure the $\ket{11}$ state probability of the two qubits. After repeating the above experiment with different modulation frequency $\nu$, we can obtain a chevron oscillation pattern of the two qubits $\ket{11}$ state probability as a function of both modulation frequency and time duration, from which the effective coupling strength $\tilde{g}$ and the resonant transition frequency between $\ket{11}$ and $\ket{02}$ can be extracted. 

In order to realize the geometric path evolution in \{$\ket{11}$, $\ket{02}$\} subspace, we apply two sinusoidal modulation drives in series and each one has a duration of $\pi/\tilde{g}$, corresponding to a $\pi$ rotation in \{$\ket{11}$, $\ket{02}$\} subspace. Therefore, in the two-qubit computational subspace \{$\ket{00}$, $\ket{01}$, $\ket{10}$, $\ket{11}$\}, the resulted unitary gate can be expressed as ~\cite{DiCarloCZ}
\begin{equation}
\label{equ:GeoCZphaseU01}
U = \left(  
\begin{array}{cccc}
1	&	0	&	0	&	0	\\
0	&	e^{i\phi_{01}}	&	0	&	0	\\
0	&	0	&	e^{i\phi_{10}}	&	0	\\
0	&	0	&	0	&	e^{i\phi_{11}}
\end{array}	
\right),
\end{equation} 
where $ \phi_{01} $  and $ \phi_{10} $ are the single-qubit phases, and $\gamma =  \phi_{11} - \phi_{01} - \phi_{10} $ is the two-qubit entangled phase.

\begin{figure}[hbp]
	\includegraphics{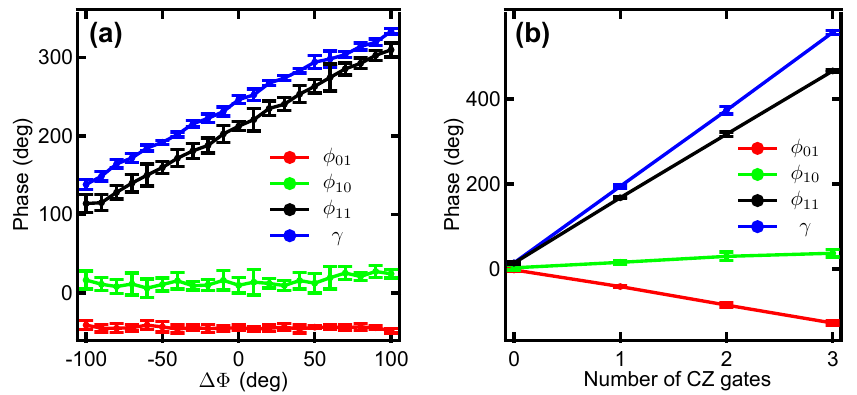}
	\caption{Calibration of the two-qubit geometric CZ gate. (a) Single- and two-qubit phases of the two-qubit state as a function of the relative phase $\Delta \Phi$ of the two sinusoidal modulation drives. (b) Single- and two-qubit phases of the two-qubit state as a function of the number of CZ gates. }
	\label{fig:GeoCZSweep}
\end{figure}

We characterize these single- and two-qubit phases by firstly initializing these two qubits in a product state $\ket{\psi_0} = \left( \ket{00} + \ket{01} + \ket{10} + \ket{11}\right)/2$, then performing two sinusoidal modulation drives in series with a relative phase $\Delta \Phi$ between them, and finally implementing the two-qubit state tomography to extract all these phases. The measurement results shown in Fig.~\ref{fig:GeoCZSweep}(a) indicate that the two-qubit entangled phase $ \gamma $ is linearly related to the relative phase $\Delta \Phi$, while the single-qubit phases $ \phi_{01} $ and $ \phi_{10} $ nearly remain the same. By choosing appropriate $\Delta \Phi$, we thus realize a two-qubit CZ gate with the entangled phase $\gamma = \pi$. The single-qubit phases can be compensated by rotating the reference frame in software. 

We further verify the realized two-qubit CZ gate by performing different numbers of CZ gates, with the experimental result shown in Fig.~\ref{fig:GeoCZSweep}(b). The measured single- and two-qubit phases are linearly related to the number of CZ gates. The small and finite offsets of the linear fits of these phases mainly come from the non-negligible crosstalk during the state preparation and measurement.

\section{Crosstalk-induced error}
During the two-qubit RB experiments, the two-qubit ZZ crosstalk will induce non-negligible errors of the single-qubit gates. Based on the two qubits\rq{} parameters in Table~\ref{Table:parameters}, we can estimate the phase shift rate of the $\ket{11}$ state caused by the ZZ crosstalk with ~\cite{Barends2014Superconducting}
\begin{equation}\label{OmegaZZ}
\Omega_{ZZ}=\frac{2g_{AB}^2 (\alpha_A+\alpha_B )}{(\omega_A - \omega_B-\alpha_A)(\omega_A - \omega_B+\alpha_B)} = -1.18~ \text{MHz},
\end{equation}
where $\omega_A$ and $\omega_B$ are the qubit frequencies of $Q_A$ and $Q_B$ at the operating spot, respectively. Thus this crosstalk interaction will induce an additional $\sim20$~\dg phase shift of the $\ket{11}$ state for an average single-qubit gate time of 40~ns, which vastly suppresses the sequence fidelity decay rates $p_\mathrm{CZ}$ and $p_\mathrm{ref}$ of the two-qubit RB experiments.

\begin{figure}[hbp]
	\includegraphics{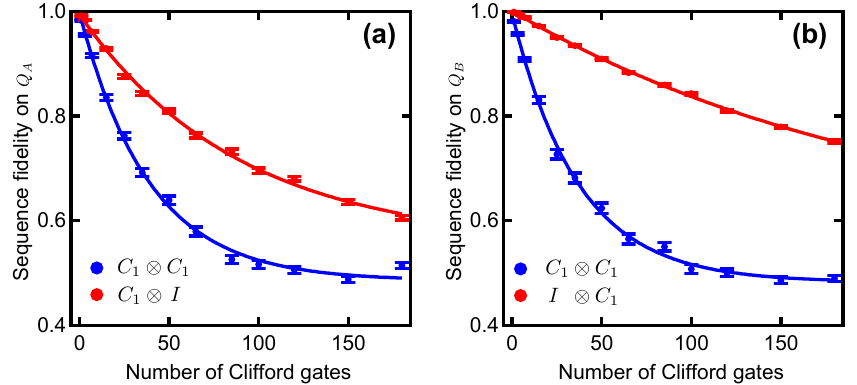}
	\caption{Simultaneous RB experimental results of qubits $Q_A$ and $Q_B$. (a) Benchmarking the crosstalk effect on qubit $Q_A$ with~(blue markers) and without~(red markers) controlling $Q_B$ simultaneously. From fitting to both decay curves, we obtain the average single-qubit gate errors $r_\mathrm{Q_A} = 0.0030(1)$ when $Q_B$ remains at the ground state and $r_\mathrm{Q_A | Q_B} = 0.0067(2)$ when $Q_B$ is also under RB experiment. (b) Benchmarking the crosstalk effect on qubit $Q_B$ with~(blue markers) and without~(red markers) controlling $Q_A$ simultaneously. From fitting to both decay curves, we obtain the average single-qubit gate errors $r_\mathrm{Q_B} = 0.0013(1)$ when $Q_A$ remains at the ground state and $r_\mathrm{Q_B | Q_A} = 0.0071(2)$ when $Q_A$ is also under RB experiment. }
	\label{fig:SimulRB}
\end{figure}

To further demonstrate the impact of ZZ crosstalk error in our system, we perform the simultaneous RB experiments for both qubits $Q_A$ and $Q_B$, in which we perform RB experiment on each qubit individually and operate both qubits simultaneously~\cite{SimultaneousRB}. The experimental results are shown in Fig.~\ref{fig:SimulRB}. The crosstalk effect on $Q_A$ when operating $Q_B$ can be determined by firstly performing RB of $Q_A$ alone~[red markers in Fig.~\ref{fig:SimulRB}(a)], then benchmarking both qubits simultaneously and tracing out the contribution of $Q_B$~[blue markers in Fig.~\ref{fig:SimulRB}(a)]. The difference of these two decay curves corresponds to the crosstalk-induced error on $Q_A$ by controlling $Q_B$. Similarly, we also find the crosstalk effect on $Q_B$ when operating $Q_A$, with the experimental result shown in Fig.~\ref{fig:SimulRB}(b). These crosstalk-induced errors have a great impact on the two-qubit RB experiments in the main text. Therefore, a tunable coupler that can switch off the coupling between adjacent qubits is desirable.


%